\documentclass[twocolumn,showpacs,preprintnumbers,amsmath,amssymb]{revtex4}

\usepackage{graphicx}
\usepackage{dcolumn}
\usepackage{bm}
\usepackage{epsfig}

\begin{document}


\title{Collisional stability of a three-component degenerate Fermi gas} 

\author{T. B. Ottenstein}
\email{timo.ottenstein@mpi-hd.mpg.de}
\author{T. Lompe}
\author{M. Kohnen}
\author{A. N. Wenz}
\author{S. Jochim}
\altaffiliation[Also at:]{Ruprecht-Karls-Universit\"at Heidelberg, Germany}

%
\affiliation{%
Max-Planck-Institut f\" ur Kernphysik, Saupfercheckweg 1, 
69117 Heidelberg, Germany
}

\date{\today}

\begin{abstract}
We report on the creation of a degenerate Fermi gas consisting of a balanced mixture of atoms in three different hyperfine states of $^6$Li. This new system consists of three distinguishable Fermions with different and tunable interparticle scattering lengths $a_{12}$, $a_{13}$ and $a_{23}$. We are able to prepare samples containing $5 \cdot 10^4$ atoms in each state at a temperature of about $215\,$nK, which corresponds to $T/T_F \approx 0.37$. We investigated the collisional stability of the gas for magnetic fields between 0 and 600\,G and found a prominent loss feature at 130\,G. From lifetime measurements we determined three-body loss coefficients, which vary over nearly three orders of magnitude.
 

\end{abstract}

\maketitle

Since the first successful preparation of a degenerate Fermi gas of $^{40}{\rm  K}$ atoms in 1999 \cite{jila_degenerate}, tremendous progress has been made in the field. The key to this success was the possibility to tune the interactions in mixtures containing two spin components at wish using Feshbach resonances \cite{ketterle_feshbach}. Large, resonant scattering lengths facilitated evaporative cooling and led to the first observation of a strongly interacting Fermi gas \cite{thomas_strongly}. At large, positive scattering lengths, weakly bound molecules could be formed \cite{jin_molecules}. These bosonic molecules were condensed into a molecular Bose-Einstein condensate (BEC) \cite{jin_bec,jochim_bec,zwierlein_bec}. The tunability of the scattering length across the Fesh\-bach resonance gave experimentalists access to the so-called BEC-BCS crossover \cite{jin_becbcs,bartenstein_becbcs,zwierlein_becbcs}, which connects in a continuous, isentropic way molecular Bose-Einstein condensates with two-component, BCS-type Fermi gases. In  this limit the  attractive interaction causes the atoms to form weakly bound many-body pairs \cite{chin_pairinggap}. In the intermediate range, where the scattering length diverges, the so-called unitary regime could be accessed \cite{thomas_unitary}, where the interactions depend only on the inverse Fermi momentum $1 /k_F$, which causes the physics to become independent of any length scale and therefore universal for any Fermi system with infinite scattering length. Consequently, such tunable two-component ultracold systems can be an ideal model for such diverse systems as neutron stars \cite{baker_neutronstars} or, with the addition of an optical lattice \cite{hofstetter_dwave}, high-$T_C$-superconductors \cite{levin_hightc}. 

\begin{figure} [htbp]
\centering
	\includegraphics [width= 8.5cm] {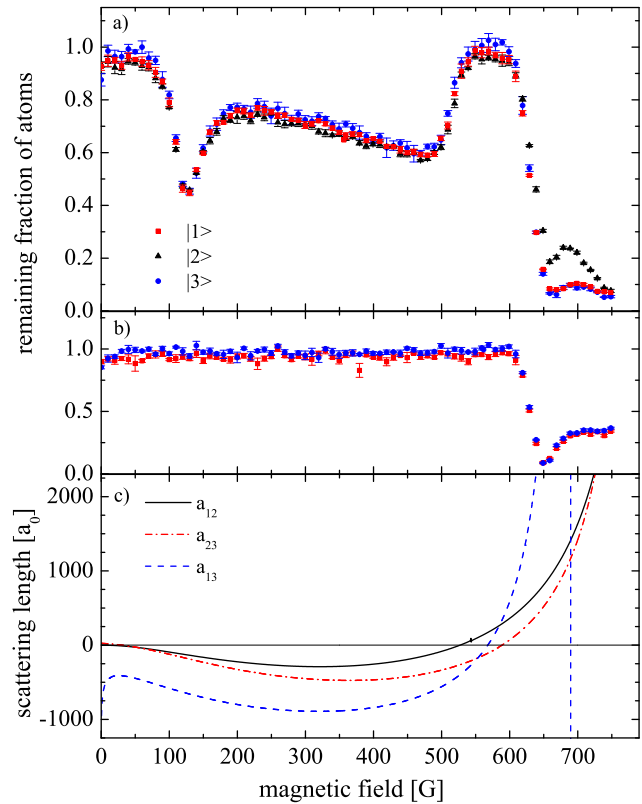}   
	
	\caption{a) Magnetic field dependence of the remaining fraction of atoms after holding a three-state mixture for 250\,ms. b) Result for a two-state mixture under the same experimental conditions. The enhanced trap loss at fields larger than 600\,G is due to high inelastic collision rates close to the two-body Fesh\-bach resonances, which also causes loss in the three-component mixture. For magnetic fields larger 600\,G, molecules are formed in the $\vert 1\rangle$ - $\vert 3\rangle$ channel which are not detected when ramping back for imaging.  c) Two-body scattering lengths for all spin-state combinations \cite{julienne} ($a_{ii}=0$  for symmetry reasons).}
\label{fig:scan0-700}
	
\end{figure}

When a third spin state is added to such a two-component gas, a number of intriguing questions arises immediately, which have already been extensively studied in theory \cite{rapp,zhai,bedaque}. How will pairing occur in such a system: Will the individual components compete, and only two of them form pairs, while the third component remains a spectator, or will the lowest-energy state of the system be a three-body bound state \cite{paananen_pairing,honerkamp,paananen_shell}? There are predictions for a phase transition between a superfluid and trionic phase in optical lattices, which can be treated analogous to baryon formation in QCD \cite{rapp_new}.
Furthermore, as the Fermi pressure is lower than in a two-component mixture of equal density, the stability of such a gas in the case of resonant two-particle interactions is still a controversial topic \cite{greene_stability,demler}. Ultracold ensembles of $^{6}{\rm  Li}$ offer the unique possibility to study these phenomena, as wide, overlapping Fesh\-bach resonances between the three lowest-lying hyperfine states allow to tune the two-body scattering lengths over a wide range (see Fig.\,\ref{fig:scan0-700}\,c)).


So far there has been only little experimental work on three-component Fermi gases: A third non-degenerate component was used as a probe for thermometry \cite{regal}. In other experiments radio frequency (RF) transitions to a third state were used to study pairing in a two-component system \cite{schunck-2008,chin_pairinggap}, but a stable three-component mixture has not been observed.  
Here we present the first realization of a stable and balanced three-component Fermi gas, which is a first step towards exploring the rich phase diagram of this system.
Furthermore, we investigate the magnetic field dependence of the collisional stability of this gas.


\begin{figure} [htbp]
\centering
	\includegraphics [width= 8cm] {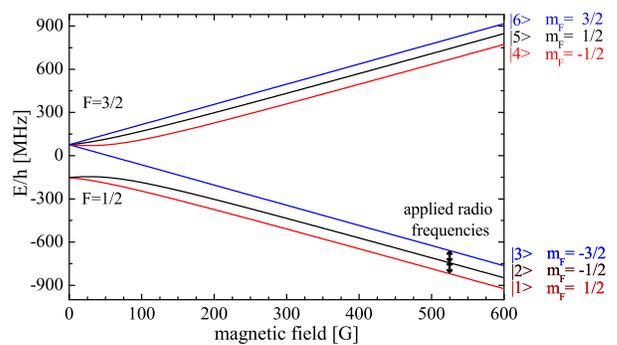}   
	\caption{Zeeman hyperfine levels of $^6Li$ in the electronic ground state. Transitions between
adjacent hyperfine states can be driven with external RF fields.}
	\label{fig:li_states}
\end{figure}

Our approach to the preparation of three-component Fermi mixtures makes use of well-established techniques used for studying ensembles of ultracold atoms \cite{thomas_alloptical}. We first precool $^6$Li atoms in a magneto-optical trap (MOT) and transfer about $10^6$ atoms into a far red-detuned optical dipole trap. During the transfer the atoms are optically pumped into the $F=1/2$ hyperfine state so that at this stage of the experiment only the two lowest magnetic subleves (labeled $\vert 1 \rangle$ and $\vert 2 \rangle$ in Fig.\,\ref{fig:li_states}) are occupied. The dipole trap is formed by crossing two counterpropagating beams with a waist of 50 $\mu$m, which results in a cigar shaped trap with an aspect ratio of about $10 : 1$. 

The wide Fesh\-bach resonance in $^6$Li allows us to tune the two-body s-wave scattering length $a_{12}$ between states $\vert 1 \rangle$ and $\vert 2 \rangle$ to $3560 \, a_0$ at a magnetic field of 751\,G, which leads to high thermalization rates. This enables us to perform fast and efficient evaporative cooling by lowering the laser power in the trap. The large positive value of the scattering length is associated with a bound state \cite{landau}, whose binding energy depends on the magnetic field. At 751\,G the binding energy is $k_B\cdot 2.3 \mu$K, where $k_B$ is Boltzmann's constant. As soon as the temperature of the sample becomes comparable to this binding energy the bound state is populated through three-body recombination. By further evaporation these molecules can be condensed into a Bose-Einstein Condensate containing $1.5\cdot 10^5$ molecules.


The third spin state $\left(\vert 3 \rangle\right)$ can be populated by driving RF transitions between the different magnetic sublevels, which has been extensively used to perform RF spectroscopy on two-component Fermi gases. In those experiments, which were performed close to the Fesh\-bach resonances, high inelastic loss prevented the preparation of stable three-component mixtures in $^6{\rm Li}$ \cite{schunck-2008}. 
We overcome the problem of loss during preparation of the mixture by tuning the magnetic field close to the zero-crossings of the two-particle scattering lengths. To avoid populating the molecular state during evaporation, we tune the scattering length to a small negative value of $a_{12} \approx -300\,a_0$ at B = 300\,G before molecule formation sets in.  There we continue evaporative cooling of the two-component mixture, reaching a temperature of 130\,nK ($T/T_F \approx 0.28$). We then minimize the interactions by ramping the magnetic field to a value of 563\,G. 




In order to prepare an equal mixture of atoms in states $\vert 1 \rangle$, $\vert 2 \rangle$ and $\vert 3 \rangle$ we simultaneously drive the RF transitions  $\vert 1 \rangle \leftrightarrow \vert 2 \rangle$ and $\vert 2 \rangle \leftrightarrow \vert 3 \rangle$ for a period of 850\,ms at a magnetic field of 563\,G. A combination of collisions and a residual magnetic field gradient lead to an incoherent mixture of atoms in the three lowest spin states. We can produce samples containing $ 5 \cdot 10^4$ atoms per spin state at a temperature of 215 nK in a harmonic trap with trap frequencies $\omega_x = \omega_y= 2 \pi \cdot  386(15) \, {\rm Hz}$ and $\omega_z= 2 \pi \cdot 38(2) \, {\rm Hz}$. Therefore, the starting point for our experiments is a degenerate three-component Fermi gas with $T/T_F \approx 0.37$ and a peak density of $ 6 \cdot 10^{11} \, {\rm atoms/cm^3}$. Under these conditions the gas is stable with a 1/e lifetime greater than 30 seconds.

In our first experiment we studied the magnetic field dependence of the collisional stability.  We prepared a three-component mixture and held the atoms at various magnetic fields between 0 and 750\,G for 250\,ms. To record the remaining fraction of atoms in one of the spin states selectively, we make use of their Zeeman splitting. We calibrated the imaging laser frequencies to the respective transitions at a magnetic field of 526\,G. Therefore we jump to this field at the end of each experimental cycle and record the number of atoms in one spin state via absorption imaging.
The result is shown in Fig.\,\ref{fig:scan0-700} a). Each data point is the average of 5 individual measurements.



In the region from 540\,G to 590\,G, where all two-particle scattering lengths cross zero, the three-component mixture is stable and the remaining fraction is close to one. For larger magnetic fields one observes a strong inelastic decay of the sample due to the two-body Fesh\-bach resonances. Below 540\,G the fraction of remaining atoms quickly drops to a minimum of 0.6 at 470\,G and then increases linearly to 0.75 at 200\,G. Between 80\,G and 190\,G there is a strong loss feature where the fraction of remaining atoms drops to about 0.5 at 130\,G. For magnetic fields below 70\,G, where $a_{12}$ and $a_{23}$ are smaller than 60 $a_0$, the mixture is once again stable.

In order to separate processes involving only two distinguishable Fermions, we did reference measurements for all possible two-state mixtures by preparing three-component mixtures and removing one component with a resonant laser pulse before ramping to the probe field. As an example, the result for the mixture of atoms in states $\vert 1\rangle$ and $\vert 3\rangle$ is shown in Fig.\,\ref{fig:scan0-700} b). Away from the Fesh\-bach resonances all two-component mixtures are stable due to Pauli blocking, as a three-body loss process would involve at least two identical Fermions. Adding a third distinguishable Fermion allows three-body processes, which can decrease the stability of the gas.  From this we infer that all decay observed in Fig.\,\ref{fig:scan0-700} a) for magnetic fields below 590\,G stems from processes involving atoms in all three spin states. 

\begin{figure} [htbp]
\centering
	\includegraphics [width= 8.5cm] {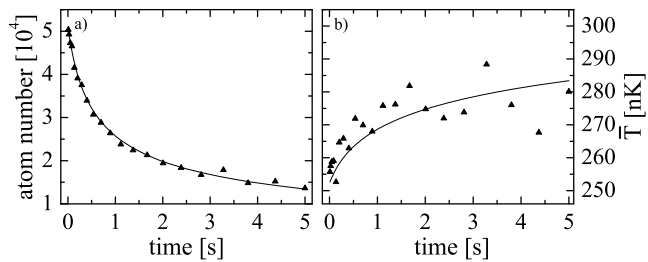}   
	\caption{Evolution of a) atom number and b) temperature  of atoms in state $\vert 2\rangle$ over five seconds at 300\,G. $\bar{T}$ denotes an effective temperature deduced from a gaussian fit to the cloud after time of flight. Each data point is a mean value of three independent measurements taken in random order. The solid line is a fit to the data, applying the method described in the text. }
\label{fig:decay_curve}
	\end{figure}
\begin{figure} [htbp]
\centering
	\includegraphics [width= 8.5cm] {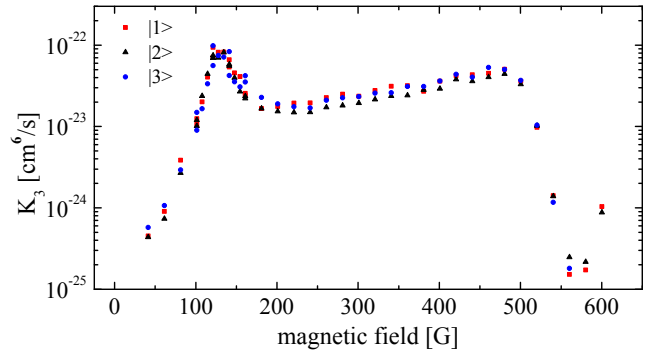}   
	\caption{Three-body loss coefficient $K_3$ vs. magnetic field.}
	\label{fig:K_3}
\end{figure}

To get quantitative information about the loss rates we measured the decay of the gas over a period of five seconds at several magnetic field values and hence for different two-body scattering lengths. For these measurements we prepared the three-component mixture as described above, tuned the magnetic field to the value of interest and measured the number and temperature of remaining atoms as a function of time. As an example, Fig.\,\ref{fig:decay_curve} shows the decay curve for state $\vert 2 \rangle$ at a magnetic field of 300\,G.
We observed that the ratio of particles in the three states remains one during the whole decay.
Therefore we can assume in the following analysis that all states have equal loss rates, that are governed by a three-body process.
The loss of particles can then be described by
\begin{equation}
\dot{n}_{i}(\vec{r})=-K_{3}\,n_{i}(\vec{r})^3,
\label{eq:local_loss_rate}
\end{equation}
where $K_{3}$ is the three-body loss coefficient and $n_i(\vec{r})$ denotes the local, temperature dependent density of atoms per spin state $\vert i \rangle$. Additionally, we take into account a small one-body loss with a $1/e$-lifetime of about 100\,s. From the two-component measurements we infer that we can neglect  two-body decay processes. To deduce $K_3$ from our experimental data, we follow a method described in \cite{innsbruck}. Loss of atoms occurs predominantly in the center of the trap where the density is high. The energy taken away by a lost particle is thus smaller than the mean energy per particle when averaging over the whole cloud. This leads to heating of the sample (see. Fig.\,\ref{fig:decay_curve} b) which requires a numerical treatment of the atom number and temperature evolution, for which we use the code developed for \cite{innsbruck}. 
The model we use describes loss in a thermal gas. For $T/T_F \approx 0.37$ minor changes occur due to degeneracy, which we neglect for simplicity. For the analysis of the lifetime curves we use an effective temperature ($\bar{T}$ in Fig.\,\ref{fig:decay_curve}) deduced from a gaussian fit to the density distribution after time of flight. If the gas is degenerate, the obtained value $\bar{T}$ is slightly higher than the real temperature. As the density in a degenerate Fermi gas is reduced with respect to a thermal gas of the same temperature, the higher temperature value compensates to some extent for the small effects of degeneracy. Hence, we approximate $T\approx\bar{T}$ for our analysis. For most values of  $K_3$ this approximation affects only the very first data points, as the temperature quickly exceeds the Fermi temperature after initial loss and heating. 
The solid line in Fig.\,\ref{fig:decay_curve} shows the fit to the atom number and temperature evolution according to this model for one decay curve. From such fits we obtain our values of $K_3$.
Fig.\,\ref{fig:K_3} shows the obtained three-body loss coefficients for all three species as a function of the magnetic field. 
As expected, $K_3$ reflects the qualitative behavior of the atom number in Fig.\,\ref{fig:scan0-700}. 
The relative error of the data points is caused by the uncertainty in spin balance, difference in detection efficiency for the three states and fitting uncertainties. This error can be estimated by the scatter of the fitted values for $K_3$  for different states and is much smaller than the observed variation of $K_3$. The absolute scale is subject to additional systematic errors in particle number and trap frequency. As $K_3\propto N^2\bar{\omega}^6$ uncertainties of $40\%$ in $N$ and $7\%$ in $\bar{\omega}$ lead to an error of $90\%$ in the absolute scale.



Although there is currently no theoretical model which can quantitatively explain the strong variation of $K_3$ with the magnetic field, we can qualitatively discuss some aspects of its behavior. 
In a naive picture one can describe a three-body event by elementary two-particle interactions governed by $a_{12}$, $a_{13}$ and $a_{23}$. If two of the three scattering lengths are close to zero, only two of the three spin states are interacting while one remains a spectator. Then the physics is comparable to a two-component system and the gas is stable.
If only one scattering length $a_{ij}$ is small, interaction between states $\vert i\rangle$ and $\vert j\rangle$ can be mediated by the third particle, allowing three-body processes.
This is consistent with our observation that in the regions below 70\,G, where $a_{12}$ and $a_{23}$ are close to zero and around 570\,G, close to the zero-crossings of $a_{13}$ and $a_{23}$, $K_3$ becomes minimal.
The smallest observed value of $10^{-25}$\,cm$^6$/s is still four orders of magnitude larger than the value for three-body recombination in a thermal cloud of $^{87}\text{\rm{Rb}}$ \cite{cornell_RbK3}, but still sufficiently small to observe lifetimes on the order of 30\,s at our densities.

At the loss feature near 130\,G the values of $K_3$ are enhanced by three orders of magnitude compared to the minimal value. In this magnetic field range all two-body scattering lengths are negative, consequently no weakly bound dimer state exists in this field range. 
As all three scattering lengths are large compared to the effective range of the interatomic potential, one would expect its short-range properties not to be relevant and therefore the physics to be universal.
If this were the case the obvious explanation for the loss feature would be a trimer state crossing into the continuum, although we cannot exclude other possible scenarios such as relaxation into deeply bound dimer states or spin changing collisions. The properties of such trimers would be comparable to Efimov trimers \cite{efimov}, which were observed through an enhanced three-body loss rate in bosonic Cs atoms \cite{innsbruck_efimov}.
An essential difference is that our system consists of distinguishable Fermions, therefore the formation of such trimers would have almost the same symmetry as the formation of baryons from three quarks in QCD.
Further theoretical and experimental investigation will be necessary to quantitatively understand the magnetic field dependence of the three-body decay rate.

In this work we have established a method to create a stable degenerate Fermi gas containing atoms in three different spin states of $^6$Li. We have determined the three-body decay rate of the gas for magnetic fields below 600\,G, whose behavior we can 
in part  explain with the two-body scattering lengths between the atoms. 
The most prominent observation is a broad loss feature caused by a process involving atoms in all three spin states, which indicates the presence of a three-particle resonance phenomenon.
The preparation of a stable three-component gas in thermal equilibrium opens up new possibilities to study many-body phenomena in ultracold gases. As an example, one could switch the scattering lengths rapidly to large values and observe the consequent collective evolution of the cloud, which would reflect the equation of state of the many-body system. At sufficiently low density, the collective time evolution should occur at a faster time scale than the loss of atoms from the trap.




We thank T. Weber and R. Grimm for providing the code used for analysis of the lifetime curves.  F. Serwane and G. Z\"urn contributed significantly to the experimental setup. Furthermore, we would like to thank P.S. Julienne, C. Chin and C.H. Greene for stimulating discussions and comments. We are grateful to J. Ullrich and his group for their generous support.

Recently we have learned that in the group of K.M.~ O'Hara similar measurements have been performed independently \cite{Ohara}.


\bibliography{paper}

\begin{thebibliography}{34}
\expandafter\ifx\csname natexlab\endcsname\relax\def\natexlab#1{#1}\fi
\expandafter\ifx\csname bibnamefont\endcsname\relax
  \def\bibnamefont#1{#1}\fi
\expandafter\ifx\csname bibfnamefont\endcsname\relax
  \def\bibfnamefont#1{#1}\fi
\expandafter\ifx\csname citenamefont\endcsname\relax
  \def\citenamefont#1{#1}\fi
\expandafter\ifx\csname url\endcsname\relax
  \def\url#1{\texttt{#1}}\fi
\expandafter\ifx\csname urlprefix\endcsname\relax\def\urlprefix{URL }\fi
\providecommand{\bibinfo}[2]{#2}
\providecommand{\eprint}[2][]{\url{#2}}

\bibitem[{\citenamefont{DeMarco and Jin}(1999)}]{jila_degenerate}
\bibinfo{author}{\bibfnamefont{B.}~\bibnamefont{DeMarco}} \bibnamefont{and}
  \bibinfo{author}{\bibfnamefont{D.~S.} \bibnamefont{Jin}},
  \bibinfo{journal}{Science} \textbf{\bibinfo{volume}{285}},
  \bibinfo{pages}{1703} (\bibinfo{year}{1999}).

\bibitem[{ket(1999)}]{ketterle_feshbach}
\emph{\bibinfo{title}{\rm{J. Stenger \textit{et al.}, Phys. Rev. Lett.
  \textbf{82}, 2422}}} (\bibinfo{year}{1999}).

\bibitem[{tho(2002{\natexlab{a}})}]{thomas_strongly}
\emph{\bibinfo{title}{\rm{K. M. O'Hara \textit{et al.}, Science \textbf{298},
  2179}}} (\bibinfo{year}{2002}{\natexlab{a}}).

\bibitem[{jin(2003)}]{jin_molecules}
\emph{\bibinfo{title}{\rm{C. A. Regal \textit{et al.}, Nature \textbf{424},
  47}}} (\bibinfo{year}{2003}).

\bibitem[{\citenamefont{Greiner et~al.}(2003)\citenamefont{Greiner, Regal, and
  Jin}}]{jin_bec}
\bibinfo{author}{\bibfnamefont{M.}~\bibnamefont{Greiner}},
  \bibinfo{author}{\bibfnamefont{C.~A.} \bibnamefont{Regal}}, \bibnamefont{and}
  \bibinfo{author}{\bibfnamefont{D.~S.} \bibnamefont{Jin}},
  \bibinfo{journal}{Nature} \textbf{\bibinfo{volume}{426}},
  \bibinfo{pages}{537} (\bibinfo{year}{2003}).

\bibitem[{joc(2003)}]{jochim_bec}
\emph{\bibinfo{title}{\rm{S. Jochim \textit{et al.}, Science \textbf{302},
  2101}}} (\bibinfo{year}{2003}).

\bibitem[{zwi(2003)}]{zwierlein_bec}
\emph{\bibinfo{title}{\rm{M. W. Zwierlein \textit{et al.}, Phys. Rev. Lett.
  \textbf{91}, 250401}}} (\bibinfo{year}{2003}).

\bibitem[{\citenamefont{Regal et~al.}(2004)\citenamefont{Regal, Greiner, and
  Jin}}]{jin_becbcs}
\bibinfo{author}{\bibfnamefont{C.~A.} \bibnamefont{Regal}},
  \bibinfo{author}{\bibfnamefont{M.}~\bibnamefont{Greiner}}, \bibnamefont{and}
  \bibinfo{author}{\bibfnamefont{D.~S.} \bibnamefont{Jin}},
  \bibinfo{journal}{Phys. Rev. Lett.} \textbf{\bibinfo{volume}{92}},
  \bibinfo{pages}{040403} (\bibinfo{year}{2004}).

\bibitem[{bar(2004)}]{bartenstein_becbcs}
\emph{\bibinfo{title}{\rm{M. Bartenstein \textit{et al.}, Phys. Rev. Lett.
  \textbf{92}, 120401}}} (\bibinfo{year}{2004}).

\bibitem[{zwi(2005)}]{zwierlein_becbcs}
\emph{\bibinfo{title}{\rm{M. W. Zwierlein \textit{et al.}, Phys. Rev. Lett.
  \textbf{94}, 180401}}} (\bibinfo{year}{2005}).

\bibitem[{chi(2004)}]{chin_pairinggap}
\emph{\bibinfo{title}{\rm{C. Chin \textit{et al.}, Science \textbf{305},
  1128}}} (\bibinfo{year}{2004}).

\bibitem[{tho(2004)}]{thomas_unitary}
\emph{\bibinfo{title}{\rm{J. Kinast \textit{et al.}, Phys. Rev. Lett.
  \textbf{92}, 150402}}} (\bibinfo{year}{2004}).

\bibitem[{\citenamefont{Baker}(1999)}]{baker_neutronstars}
\bibinfo{author}{\bibfnamefont{G.~A.} \bibnamefont{Baker}},
  \bibinfo{journal}{Phys. Rev. C} \textbf{\bibinfo{volume}{60}},
  \bibinfo{pages}{054311} (\bibinfo{year}{1999}).

\bibitem[{hof(2002)}]{hofstetter_dwave}
\emph{\bibinfo{title}{\rm{W. Hofstetter \textit{et al.}, Phys. Rev. Lett.
  \textbf{89}, 220407}}} (\bibinfo{year}{2002}).

\bibitem[{lev(2005)}]{levin_hightc}
\emph{\bibinfo{title}{\rm{Q. Chen \textit{et al.}, Phys. Rep. \textbf{412},
  1}}} (\bibinfo{year}{2005}).

\bibitem[{rap(2007)}]{rapp}
\emph{\bibinfo{title}{\rm{A. Rapp \textit{et al.}, Phys. Rev. Lett.
  \textbf{98}, 160405}}} (\bibinfo{year}{2007}).

\bibitem[{\citenamefont{Zhai}(2007)}]{zhai}
\bibinfo{author}{\bibfnamefont{H.}~\bibnamefont{Zhai}}, \bibinfo{journal}{Phys.
  Rev. A} \textbf{\bibinfo{volume}{75}}, \bibinfo{eid}{031603}
  (\bibinfo{year}{2007}).

\bibitem[{\citenamefont{Bedaque and D'Incao}(2006)}]{bedaque}
\bibinfo{author}{\bibfnamefont{P.~F.} \bibnamefont{Bedaque}} \bibnamefont{and}
  \bibinfo{author}{\bibfnamefont{J.~P.} \bibnamefont{D'Incao}},
  \bibinfo{journal}{arXiv:cond-mat/0602525}  (\bibinfo{year}{2006}).

\bibitem[{\citenamefont{Paananen et~al.}(2006)\citenamefont{Paananen,
  Martikainen, and T\"{o}rm\"{a}}}]{paananen_pairing}
\bibinfo{author}{\bibfnamefont{T.}~\bibnamefont{Paananen}},
  \bibinfo{author}{\bibfnamefont{J.-P.} \bibnamefont{Martikainen}},
  \bibnamefont{and}
  \bibinfo{author}{\bibfnamefont{P.}~\bibnamefont{T\"{o}rm\"{a}}},
  \bibinfo{journal}{Phys. Rev. A} \textbf{\bibinfo{volume}{73}},
  \bibinfo{eid}{053606} (\bibinfo{year}{2006}).

\bibitem[{\citenamefont{Honerkamp and Hofstetter}(2004)}]{honerkamp}
\bibinfo{author}{\bibfnamefont{C.}~\bibnamefont{Honerkamp}} \bibnamefont{and}
  \bibinfo{author}{\bibfnamefont{W.}~\bibnamefont{Hofstetter}},
  \bibinfo{journal}{Phys. Rev. B} \textbf{\bibinfo{volume}{70}},
  \bibinfo{pages}{094521} (\bibinfo{year}{2004}).

\bibitem[{\citenamefont{Paananen et~al.}(2007)\citenamefont{Paananen,
  T\"{o}rm\"{a}, and Martikainen}}]{paananen_shell}
\bibinfo{author}{\bibfnamefont{T.}~\bibnamefont{Paananen}},
  \bibinfo{author}{\bibfnamefont{P.}~\bibnamefont{T\"{o}rm\"{a}}},
  \bibnamefont{and} \bibinfo{author}{\bibfnamefont{J.-P.}
  \bibnamefont{Martikainen}}, \bibinfo{journal}{Phys. Rev. A}
  \textbf{\bibinfo{volume}{75}}, \bibinfo{eid}{023622} (\bibinfo{year}{2007}).

\bibitem[{\citenamefont{Rapp et~al.}(2008)\citenamefont{Rapp, Hofstetter, and
  Zar\'{a}nd}}]{rapp_new}
\bibinfo{author}{\bibfnamefont{A.}~\bibnamefont{Rapp}},
  \bibinfo{author}{\bibfnamefont{W.}~\bibnamefont{Hofstetter}},
  \bibnamefont{and}
  \bibinfo{author}{\bibfnamefont{G.}~\bibnamefont{Zar\'{a}nd}},
  \bibinfo{journal}{Phys. Rev. B} \textbf{\bibinfo{volume}{77}},
  \bibinfo{eid}{144520} (\bibinfo{year}{2008}).

\bibitem[{gre(2008)}]{greene_stability}
\emph{\bibinfo{title}{\rm{D. Blume \textit{et al.}, Phys. Rev. A \textbf{77},
  033627}}} (\bibinfo{year}{2008}).

\bibitem[{\citenamefont{Cherng et~al.}(2007)\citenamefont{Cherng, Refael, and
  Demler}}]{demler}
\bibinfo{author}{\bibfnamefont{R.~W.} \bibnamefont{Cherng}},
  \bibinfo{author}{\bibfnamefont{G.}~\bibnamefont{Refael}}, \bibnamefont{and}
  \bibinfo{author}{\bibfnamefont{E.}~\bibnamefont{Demler}},
  \bibinfo{journal}{Phys. Rev. Lett.} \textbf{\bibinfo{volume}{99}},
  \bibinfo{eid}{130406} (\bibinfo{year}{2007}).

\bibitem[{\citenamefont{Regal}(2005)}]{regal}
\bibinfo{author}{\bibfnamefont{C.~A.} \bibnamefont{Regal}},
  \bibinfo{journal}{Ph.D. thesis, University of Colorado, Boulder}
  (\bibinfo{year}{2005}).

\bibitem[{sch(2008)}]{schunck-2008}
\emph{\bibinfo{title}{\rm{C. H. Schunck \textit{et al.}, Nature \textbf{454},
  739}}} (\bibinfo{year}{2008}).

\bibitem[{tho(2002{\natexlab{b}})}]{thomas_alloptical}
\emph{\bibinfo{title}{\rm{S. R. Granade \textit{et al.}, Phys. Rev. Lett.
  \textbf{88}, 120405}}} (\bibinfo{year}{2002}{\natexlab{b}}).

\bibitem[{\citenamefont{Landau and Lifshitz}(1981)}]{landau}
\bibinfo{author}{\bibfnamefont{L.~D.} \bibnamefont{Landau}} \bibnamefont{and}
  \bibinfo{author}{\bibfnamefont{L.~M.} \bibnamefont{Lifshitz}},
  \emph{\bibinfo{title}{Quantum Mechanics}} (\bibinfo{publisher}{Butterworth
  Heinemann}, \bibinfo{year}{1981}), \bibinfo{edition}{3rd} ed.

\bibitem[{jul(2005)}]{julienne}
\emph{\bibinfo{title}{\rm{P. S. Julienne (private communication)}}},
  \bibinfo{address}{calculated using the model described in M. Bartenstein
  \textit{et al.}, Phys. Rev. Lett. \textbf{94}, 103201}
  (\bibinfo{year}{2005}).

\bibitem[{inn(2003)}]{innsbruck}
\emph{\bibinfo{title}{\rm{T.Weber \textit{et al.}, Phys. Rev. Lett.
  \textbf{91}, 123201}}} (\bibinfo{year}{2003}).

\bibitem[{cor(1997)}]{cornell_RbK3}
\emph{\bibinfo{title}{\rm{E. A. Burt \textit{et al.}, Phys. Rev. Lett.
  \textbf{79}, 337}}} (\bibinfo{year}{1997}).

\bibitem[{\citenamefont{Efimov}(1971)}]{efimov}
\bibinfo{author}{\bibfnamefont{V.}~\bibnamefont{Efimov}},
  \bibinfo{journal}{Sov. J. Nucl. Phys.} \textbf{\bibinfo{volume}{12}},
  \bibinfo{pages}{589} (\bibinfo{year}{1971}).

\bibitem[{inn(2006)}]{innsbruck_efimov}
\emph{\bibinfo{title}{\rm{T. Kraemer \textit{et al.}, Nature \textbf{440},
  315}}} (\bibinfo{year}{2006}).

\bibitem[{Oha(2008)}]{Ohara}
\emph{\bibinfo{title}{\rm{J. H. Huckans \textit{et al.}, arXiv:0810.3288}}}
  (\bibinfo{year}{2008}).

\end{thebibliography}

\end{document}